\documentclass[aps,twocolumn,superscriptaddress]{revtex4}
\usepackage[T1]{fontenc}
\usepackage{amsmath} 
\usepackage{mathtools}
\usepackage{physics} 
\usepackage{amsfonts} 
\usepackage{booktabs} 
\usepackage{graphicx} 
\usepackage{float} 
\usepackage{multirow}
\usepackage[caption=false]{subfig}
\usepackage{hyperref}
\usepackage{xcolor}
\usepackage{upgreek}
\usepackage{siunitx}
\usepackage{verbatim}
\usepackage{ulem}

\begin{document}

\title{A new method for spatial mode shifting of a stabilized optical cavity for the generation of dual-color X-rays}

\author{Edoardo~Suerra}
\affiliation{Universit\`a degli Studi di Milano - Via Celoria, 16, 20133 Milano, Italy}
\affiliation{Istituto Nazionale di Fisica Nucleare - Sezione di Milano - Via Celoria, 16, 20133 Milano, Italy}

\author{Dario~Giannotti}
\affiliation{Istituto Nazionale di Fisica Nucleare - Sezione di Milano - Via Celoria, 16, 20133 Milano, Italy}
 
\author{Francesco~Canella}
\affiliation{Politecnico di Milano - Piazza Leonardo da Vinci, 32, 20133 Milano, Italy}
\affiliation{Istituto Nazionale di Fisica Nucleare - Sezione di Milano - Via Celoria, 16, 20133 Milano, Italy}

\author{Stefano~Capra}
\affiliation{Universit\`a degli Studi di Milano - Via Celoria, 16, 20133 Milano, Italy}

\author{Daniele~Cipriani}
\affiliation{Universit\`a degli Studi di Milano - Via Celoria, 16, 20133 Milano, Italy}
\affiliation{Istituto Nazionale di Fisica Nucleare - Sezione di Milano - Via Celoria, 16, 20133 Milano, Italy}

\author{Giovanni~Mettivier}
\affiliation{Universit\`a degli Studi di Napoli Federico II, Via Cintia, 21, 80126 Napoli, Italy}
\affiliation{Istituto Nazionale di Fisica Nucleare - Sezione di Napoli - Via Cintia, 21, 80126 Napoli, Italy}

\author{Gianluca~Galzerano}
\affiliation{Istituto di Fotonica e Nanotecnologie - Consiglio Nazionale delle Ricerche, Piazza Leonardo da Vinci, 32, 20133 Milano, Italy}
\affiliation{Politecnico di Milano - Piazza Leonardo da Vinci, 32, 20133 Milano, Italy}

\author{Paolo~Cardarelli}
\affiliation{Istituto Nazionale di Fisica Nucleare - Sezione di Ferrara - Via Saragat, 1, 44122  Ferrara, Italy}

\author{Simone~Cialdi}
 \email{simone.cialdi@unimi.it}
\affiliation{Universit\`a degli Studi di Milano - Via Celoria, 16, 20133 Milano, Italy}
\affiliation{Istituto Nazionale di Fisica Nucleare - Sezione di Milano - Via Celoria, 16, 20133 Milano, Italy}

\author{Luca~Serafini}
\affiliation{Istituto Nazionale di Fisica Nucleare - Sezione di Milano - Via Celoria, 16, 20133 Milano, Italy}

\begin{abstract}
We propose an innovative method to shift the transversal position of the focal point of an optical
cavity keeping it actively stabilized.  Our cavity is a 4 mirrors bow-tie cavity and the spatial shift of the resonant mode
is obtained by properly rotating the two curved mirrors by piezo actuators. This method allows us to move the transversal position of the cavity focal point of \SI{135}{\micro\meter}
in a time of \SI{50}{\milli\second}, keeping the resonance condition of the cavity by means of the Pound-Drever-Hall technique.
We propose to use this technique for the generation of 2-color X-rays via Inverse Compton Scattering (ICS).
This technique exploits the large
average power stored in the high finesse cavity by shifting the laser beam with respect to the electron beam trajectory, hence controlling
the spatial superposition of the electron and photon beams in the interaction region.
Arranging two cavities assembled one on top of the other, with different collision angle with the electron beam, allows the generation of X-ray bursts of different energies just by swiftly moving the two cavities, switching the two focal points onto
the electron beam trajectory, thus activating in sequence two different ICS spectral lines.
\end{abstract}

\maketitle

\section{Introduction}
Dual-energy imaging, proposed originally in 1976 by Alvarez and Macovski \cite{Alv}, is a radiography modality that uses two images acquired with two different X-ray spectra, so that the information about the energy-dependent attenuation can be used to discriminate different materials.
In the last years, dual-energy computed tomography has demonstrated the potential to become a well-established diagnostic tool in clinical routine \cite{collough2015dual}, despite the limitations due to the broad energy bandwidth of X-ray spectra from the conventional X-ray tubes used as the radiation source.
Monochromatic X-ray sources could allow to fully take advantage of dual-energy techniques, such as K-edge digital subtraction (KES), which can be used for contrast enhancement of specific features by the administration of a suitable contrast agent. This technique is based on the K-edge discontinuity in the photoelectric absorption of the contrast element. By acquiring two images with two monochromatic beams having energies bracketing the K-edge, it is possible to isolate an image of the contrast agent distribution using subtraction algorithms. This suppresses the anatomical noise, i.e., signals of other surrounding tissues, effectively enhancing the detectability of the features of interest.

The implementation of KES requires two X-ray beams having different mean energies close to the K-edge and narrow bandwidths to avoid spectra overlap. Several experiments of K-edge subtraction imaging in the biomedical field have been carried out at synchrotron facilities worldwide, demonstrating its potential clinical applications as reviewed in the work by Thomlinson et al.  \cite{thomlinson2018k}.
Nonetheless, a translation of synchrotron radiation to clinical routine is not achievable, due to the size and cost of these infrastructures.
The research for a novel intense monochromatic X-ray source to fill the gap between  synchrotron radiation and conventional X-ray tubes is a very active field in the scientific community.  

Inverse Compton Scattering (ICS), i.e. the scattering of laser light by relativistic electrons, is currently one of the more promising solutions to provide intense tunable monochromatic X-rays by means of a compact source. 

The radiation produced by ICS sources exibits an energy-angle correlation, with the most energetic photons emitted on axis and the photon energy decreasing as the scattering angle increases.
For a laser photon with energy $E_L=h\nu$ colliding with a single electron, the energy of the backscattered photon $E$ is related to the polar scattering angle $\theta$ as:
\begin{equation}
E=\frac{2E_L\gamma^{2}\left(1+\cos\alpha\right)}{1+\gamma^{2}\theta^{2}},
\label{eq:energy}
\end{equation}
where $\gamma$ is the electron relativistic Lorentz factor and $\alpha$ is the coliision angle between the laser and the electron beam.
The energy-angle correlation permits to adjust the energy bandwidth by using irides or collimators, thus removing the radiation emitted outside of a selected acceptance angle. 
Equation (\ref{eq:energy}) holds in the case of a negligible electron recoil and low laser intensity regime. A more general and complete treatment can be found in specialised works, such as \cite{petrillo2012photon,sun2011theoretical,curatolo2017analytical}.

The advance in the laser and particle acceleration technology occurred in the last decades allowed the current ongoing transition of ICS sources from research-oriented
or demonstrative machines, towards actual user facilities.
Several laboratories and research institutions worldwide are operating or proposing projects for the development and commissioning of ICS radiation facilities.
Currently, a user facility is operational in Europe: MuCLS at the Technical University of Munich, Germany \cite{eggl2016munich,gunther2020versatile}, while other two are under commissioning: STAR (University of Calabria, Cosenza, Italy) \cite{faillace2019status,bacci2014star} and ThomX (LAL, Orsay, France) \cite{Dupraz2020Thomx}.

Within this scenario, a proposal for a Multidisciplinary Advanced Research Infrastructure with X-rays (MariX)  based on high repetition-rate energy-recovery superconductive linac has been supported by Università di Milano and the Italian Institute for Nuclear Physics (INFN) \cite{serafini2019marix,Marix_CDR}.
An ICS source is conceived as a first-phase development of the entire MariX facility: the Bright compact X-ray Source (BriXS) \cite{Cardarelli2020BriXS}. This will be a user-oriented ICS facility for hard X-rays application in the bio-medical field, as well as non-destructive testing for cultural heritage and materials science. 
A compact demonstrator of the MariX-BriXS system is under design and proposed to be installed at INFN Milan-LASA laboratory, aimed at demonstrating both the principle of two-way acceleration (see \cite{bacci2019}) and the ERL operation of a ICS with \SI{5}{\milli\ampere} average current electron beam and up to \SI{45}{\mega\electronvolt} energy. BriXsino will have a footprint of about $\SI{40}{}\times\SI{15}{m}$ (bunker size), aiming at generating Compton backscattering X-rays of up to \SI{37}{\kilo\electronvolt} energy and \SI{4e11}{photon\per\second} fluxes within $5\%$ rms relative bandwidths.

Each electron bunch will interact with the laser pulse stored in the ring cavity at the interaction points IP with a maximum repetition rate of 100 MHz.

For a continuously tunable X-ray energy, as required by the majority of medical imaging applications, the typical approach used on the ICS sources currently operating is to adjust the electron energy.
The amount of time necessary to change the electron energy and obtain a stable working point can be of the order of several minutes, as well as hours. This is obviously not compatible with in-vivo dual-energy medical imaging, which requires instead a fast switch between the two energies to avoid motion effects.

X-ray imaging tests with ICS sources using contrast agents have been reported with (quasi-)monochromatic beam at a single energy above the K-edge of iodine \cite{kuroda2014k,Eggl2017}, as well as using filtration of the same material used as contrast medium and specific reconstruction algorithms \cite{kulpe2018,kulpe2020k}.
Also, simulations of the application of KES technique with X-ray beams of proposed ICS facilities demonstrated the potential of these sources in the case of coronary angiography and contrast-enhanced dual-energy mammography \cite{Paterno2020dual,Paterno2019_kedge}.
Nonetheless, dual-energy KES experiments with ICS have never been carried out so far, due to the lack of an effective implementation of a fast dual-energy switch.

In this paper we describe a new technique capable to achieve the production of X-rays from ICS with switching times between two energies of about \SI{50}{} milliseconds. This setup, when implemented in the BriXS/BriXSino facility will make possible for the first time to perform dual-energy imaging experiments with ICS radiation.

\begin{figure}[ht]
\centering
\includegraphics[width=0.45\textwidth]{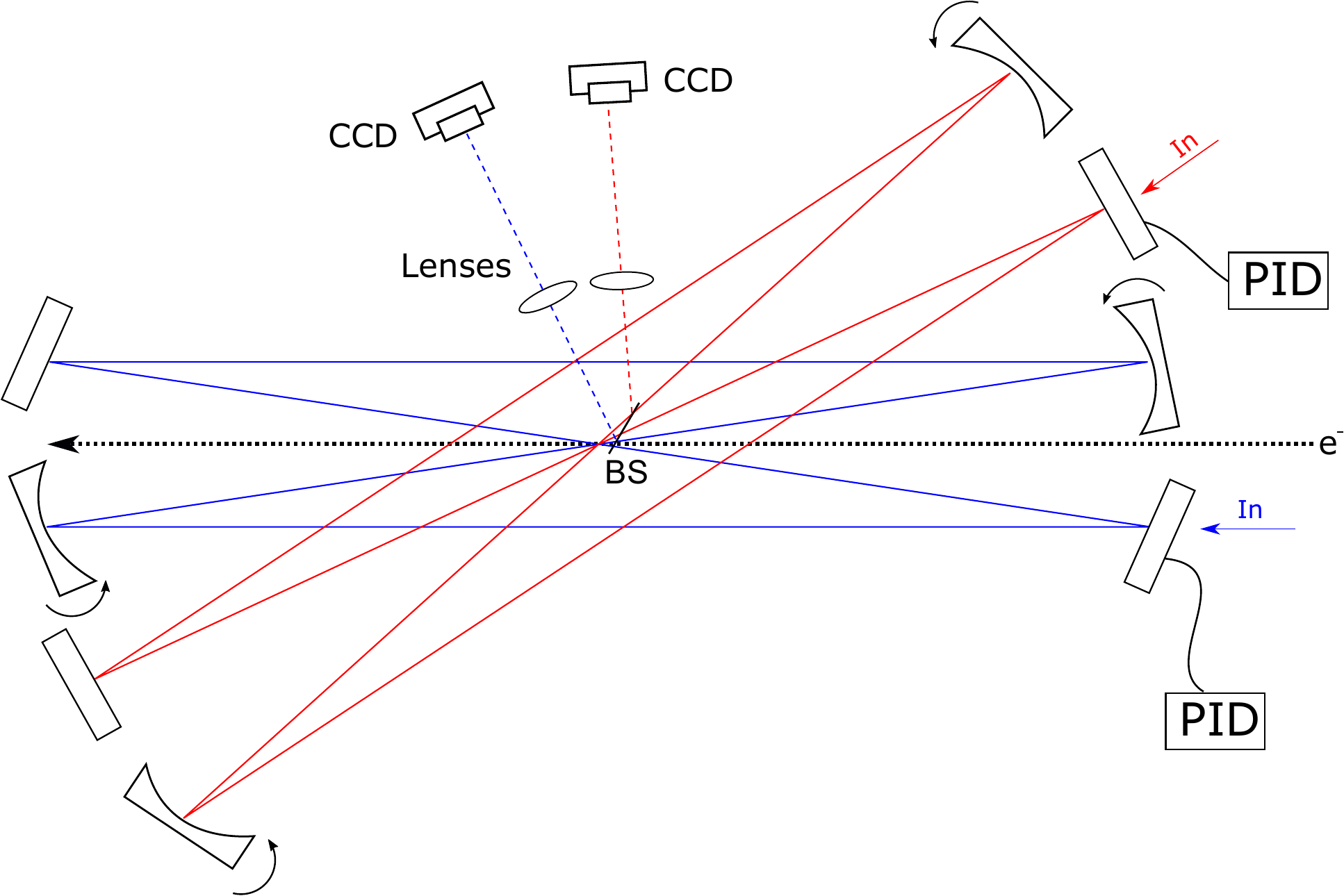}
\caption{Setup used in our work. There are two 4-mirrors bow-tie crossed cavities, each actively stabilized with respect to the same external laser. The blue cavity will be used for the generation of X-rays with higher frequency, while the red cavity with lower frequency.}
\label{fig:2cavities}
\end{figure}

As shown in Eq. (\ref{eq:energy}), to adjust the mean energy of the X-rays emission it is possible to act on three parameters: the electron energy, the laser wavelength or the collision angle.  In order to enable a dual-color switch without changing the electron energy and without the introduction of a second laser system, we plan to make the electron beam interact with two laser beams having the same wavelength, but at two different collision angles \cite{drebot2017two}. 
The switch between the two energies can be made by swapping the interaction laser. Our proposal  is based on two crossed ring cavities and a new technique to shift the intracavity focal points positions of 
$>100\mu m$ in a time $<100ms$ in order to switch between these two cavities
and in turn to change the collision angle between photons and electrons (Fig.~\ref{fig:2cavities}).

The paper is structured as follows.
In Section "Method" we describe the technique we use to move the focal point of the two cavities and we show the related simulations. 
In Section "Results" we report on the measurements about the spatial shift of the focal points and the time it takes to move them.
In particular, we show the comparison of the cavity transmissions before and after the spatial shift of the focal point.
Section "Conclusions" closes the paper with some concluding remarks.

\section{Method}
The aim of our work is to move the focal points of two ring optical cavities while they are actively frequency stabilized to an external laser.
This task serves two purposes:
\begin{enumerate}
\item the fine-tuning of the overlap of the photon beam inside a cavity over an electron bunch;
\item the implementation of the dual-color method.
\end{enumerate}
To achieve these requests, the two ring-cavity must remain in resonance with the external laser as the focal points move.

Before focusing on the method itself, it is fundamental to describe our experimental setup. It is shown in Fig.~\ref{fig:2cavities} and it is composed by two planar 4-mirrors bow-tie cavities.
For sake of simplicity we label the two cavities as \textit{Blue} and \textit{Red}: in the final ICS setup, blue cavity will generate X-rays with higher frequency, while red cavity with lower frequency.
Both cavities are injected by a \SI{1030}{\nano\meter} Yb mode-locked laser with a repetition rate of \SI{100}{\mega\hertz}.
They are constituted by two plane mirrors and two curved mirrors with \SI{750}{\milli\meter} radius of curvature, where the input coupler is plane and it is connected to a cylindric piezoelectric actuator which serves as active stabilization via Pound-Drever-Hall technique \cite{bib:pdh}.
The curved mirrors distance is \SI{755}{\milli\meter}, thus cavities are in a near-confocal configuration, corresponding to the minimum spot size in the cavity focusing point, as usually is for ICS applications.
The focal point of each cavity is in the middle point between their respective two curved mirrors and it is slightly elliptical since the incidence angle between the beam and the curved mirrors is \SI{3.5}{\degree}.
The blue and red cavities are geometrically identical and they are rotated one to the other so that the interaction angles between the photons and the electrons direction are \SI{7}{\degree} and \SI{30}{\degree}, respectively.
The Finesse of our cavities is essentially defined by the input coupler reflectivity, since the other three mirrors have a very high reflectivity ($>$\SI{99.99}{\percent}). 
The Free Spectral Range of the cavities is \SI{100}{\mega\hertz}, in resonance with the laser pulse repetition rate, and the input average power is around \SI{10}{\milli\watt} for each cavity.
Finesse is measured using the setup described in \cite{bib:galzerano} giving $F=1800$ for both cavities. This value corresponds to an input coupler reflectivity of $R_{\rm in}=\SI{99.67}{\percent}$.

\begin{figure}[ht]
\centering
\includegraphics[width=0.4\textwidth]{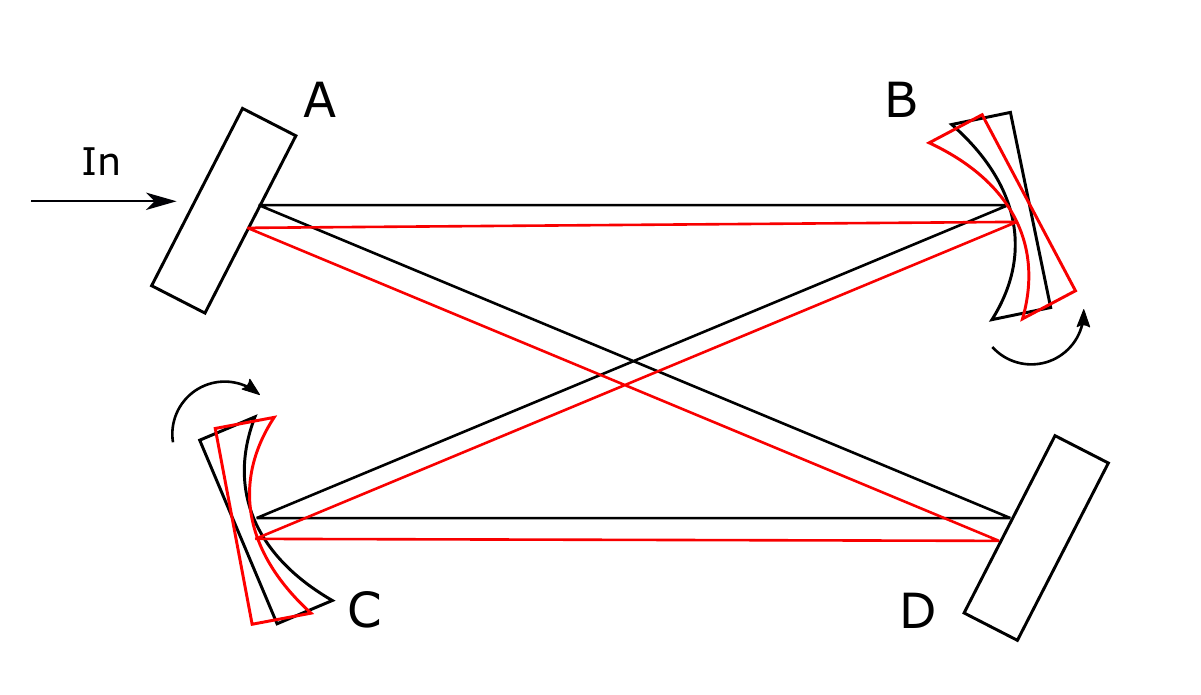}\\
\includegraphics[width=0.3\textwidth]{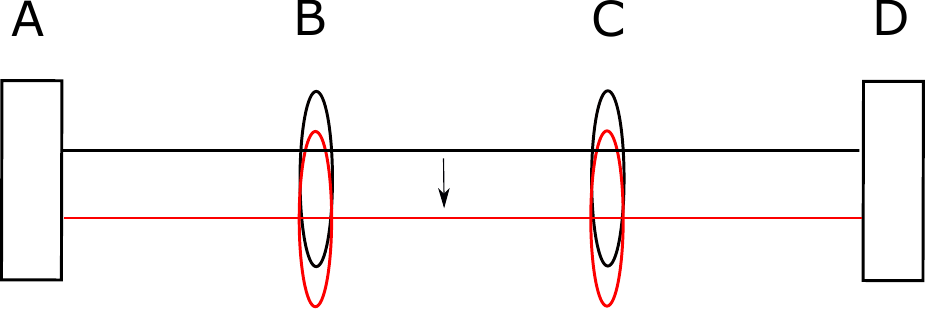}
\caption{Above: scheme of the shift method. Here the curved mirrors rotate around a vertical axis, so a horizontal shift of the focus (and of all the optical axis) is obtained. The same would happen for a rotation around a horiziontal axis, where the focus would shift vertically.
Below: equivalent $2$-mirrors system, where the curved mirrors are represented by $2$ lenses. The mirrors rotation is equivalent to a lenses shift.}
\label{fig:method}
\end{figure}

Upper part of  Fig.~\ref{fig:method}  shows the method we propose for the shift of each focus.
The core of our technique resides in the rotation of the two curved mirrors $B$ and $C$: if they are rotated by the same angle but with inverse rotation direction, then a rigid displacement of the mode inside the cavity is obtained, leaving the cavity aligned.
Moreover, when the axis of rotation intersects the point where the beam is incident on the mirror, the length of the round-trip does not change.
If this condition is not perfectly fullfilled, then the active stabilization system of the cavity will automatically compensate for the variation in the length of the round-trip.
The lower part of Fig.~\ref{fig:method} shows the equivalent scheme representing the cavity as linear with two lenses to simplify the interpretation of the method.
The effect of the rotation of the two curved mirrors is equivalent to a shift of the two lenses with consequent rigid displacement of the mode of the cavity.
Fig.~\ref{fig:method} shows an horizontal displacement of the cavity modes, since mirror rotates around a vertical axis, but a vertical displacement can be obtained equivalently by rotating mirrors around an horizontal axis.
In our work we focus on the last case, where a vertical shift of the two foci is performed.
For small rotation angles, a linear relationship is found between the shift and rotation angle of the mirrors corresponding to \SI{188}{\micro\meter\per\milli\radian} for our cavity design.

\begin{figure}[ht]
\centering
\includegraphics[width=0.45\textwidth]{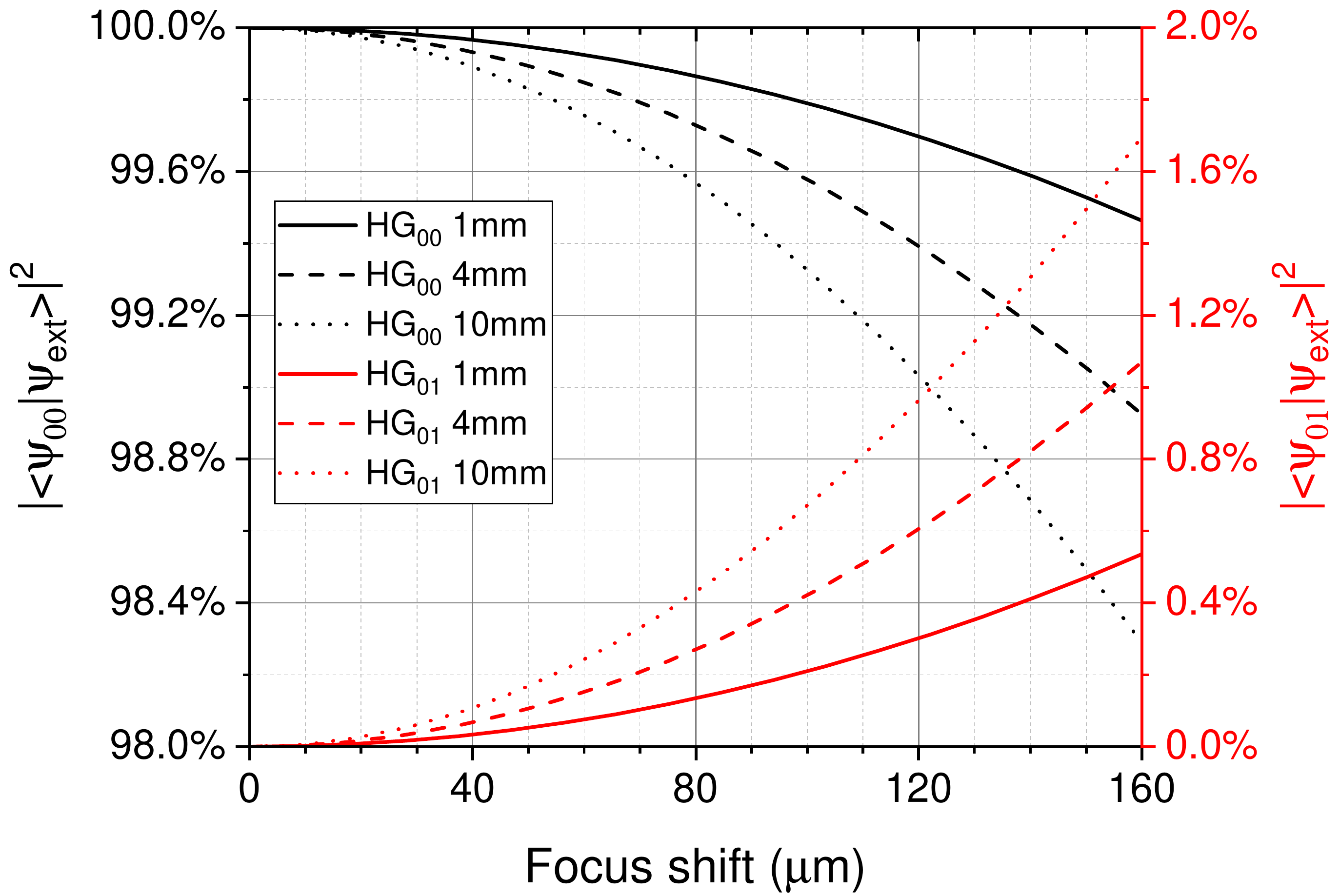}
\caption{Shift method simulations.
Black lines (left axis) show the projection of the external $\rm HG_{00}$ mode onto the internal $\rm HG_{00}$ mode in function of the focus shift for different cavity geometries: curved mirrors distance is $\SI{1}{\milli\meter}$, $\SI{4}{\milli\meter}$ and $\SI{10}{\milli\meter}$ from the confocal condition, respectively.
Red lines (right axis) show the projection of the external $\rm HG_{00}$ mode onto the internal $\rm HG_{01}$ mode for the same cavity geometries. In every configuration we assume that external and internal $\rm HG_{00}$ are identical for a \SI{0}{\micro\meter} focus shift.
Note: measurements in "Results" have been performed in the $\SI{4}{\milli\meter}$ condition (dashed lines).}
\label{fig:method-sim}
\end{figure}

While the cavity remains aligned after the mirrors rotation, it does not remain aligned with the external input mode, since the whole optical axis of the cavity rigidly translates: the non-perfect overlap with the external mode produces a loss in power inside the cavity (reduction of cavity coupling efficiency).
Indeed, if the external mode is a Hermite-Gaussian $\rm HG_{\rm 00}$, the spatial shift of the cavity optical axis produces a non-zero projection on other $\rm HG$ modes. The projection is almost completely on the $\rm HG_{\rm 01}$, while the others modes are negligible. Since $\rm HG_{\rm 01}$ has a different frequency than the fundamental mode, when $\rm HG_{\rm 00}$ is resonant then $\rm HG_{\rm 01}$ is not, so the power projected onto $\rm HG_{\rm 01}$ is reflected and does not accumulates in the cavity.
We simulate these effects for our cavity geometry and we show the results in Fig.~\ref{fig:method-sim}: here we calculated the squared projection of the external mode onto both the internal $\rm HG_{\rm 00}$ and $\rm HG_{\rm 01}$ mode as a function of the focus shift for $3$ different geometrical cavity configurations. This corresponds to the power coupling between external and internal mode.
In the simulation three configurations are shown, with different distances between the two curved mirrors and so with different distances from confocal configuration. The confocal configuration is for $BC=\SI{751.4}{\milli\meter}$, dominated by the vertical  
component of the beam, which is elliptical so that confocal distance is different in the horizontal 
and vertical direction.
The first configuration is at \SI{1}{\milli\meter} from the confocal, the second at \SI{4}{\milli\meter} and the third at \SI{10}{\milli\meter}.
In all of these configurations we assume that internal and external $\rm HG_{00}$ modes are identical for a \SI{0}{\micro\meter} focus shift.
Experimentally, this means that if we change $BC$ distance, we have to change the external mode diameter to compensate the internal mode diameter variation, since it depends on $BC$.
Simulations shows that the overlap depends on the choosen configuration, as one would expect: indeed, approaching the confocal configuration the mode on the input mirror becomes wider, reducing the relative shift of the optical axis with respect to the mode dimension and so reducing the projection of the external  $\rm HG_{00}$ onto the internal $\rm HG_{01}$ for a given focus shift.
Since we are interested in focus movements of the order of \SI{100}{\micro\meter} and our cavity is in the \SI{4}{\milli\meter} configuration of Fig.~\ref{fig:method-sim}, the misalignment produces a negligible loss in power of less than \SI{1}{\percent} in the fundamental mode.

Experimentally, every curved mirror is placed on a mounting whose rotation is remotely controlled via piezolectric actuators.
We tested two different types of mounting on the two cavities: standard mountings on the blue cavity and gimbal mountings on the red cavity.
The difference is that standard mountings can rotate around an axis lying outside the mirror and so shifted with respect to the incidence point of the beam, while the gimbal mountings can rotate around the incidence point. In the last case, the mirror rotation does not affect the cavity length, so the stabilization system can compensate more easily.

This method allows the very fine alignment of the photons with respect to the electrons, by using the following procedure.
Firstly, a pre-alignment is required.
At the center of the cavity there is an iris in the exact position of the focal point of the electron beam.
Both the laser beams entering the cavities and the cavities themselves are aligned in a way such that the mode of the cavity passes exactly to the center of the iris.
With this pre-alignment a $\sim\SI{100}{\micro\meter}$ accuracy of the focal points positions can be achieved.
This can be confirmed by placing a simple imaging system to monitor the position of the foci, as depicted in Fig.~\ref{fig:2cavities}: a pellicle beam splitter reflects part of the beams to a lens and then to a CCD camera, where the image forms.
By adjusting the position of the lens and the CCD we do the imaging of the conjugate plane of interest, where the iris and the two foci lies.
If we back-illuminate the closed iris, the right lens-CCD configuration corresponds to a sharp image of the iris on the CCD.
At this point magnification of the imaging system is measured by placing an object of known dimensions against the iris and comparing it with the corresponding measured dimensions.
We can exploit this system to overlap the two foci in the closed iris with sub-\SI{100}{\micro\meter} accuracy.
At this point we can stabilize the two cavities and apply our method for a very fine alignment: we will monitor and maximize the power of the X-rays produces by each cavity.
It is worth noting that the presence of the pellicle beam splitter does not introduce any walk off of the beam inside the cavities, maintaining them aligned, but on the other hand it significantly reduces the Finesse of the cavities.
Obviously the pellicle beam splitter can only be used to monitor the two foci during the alignment/calibration, then it must be removed for the X-rays generation.
Currently, we remove it by hand, but in the final setup a remote-controlled removing system must be used, since we will operate in vacuum.

As previously explained, our method can be exploited for the implementation of dual color method, too.
In this case the strategy is the following.
Firstly, one optimizes the overlap between photons and electrons for both cavities, then the two foci are shifted in order to set the X-rays power to zero.
At time $t=\SI{0}{\second}$ the focus of the blue cavity is shifted so that its photons are overlapped to the electrons and this configuration is maintained for \SI{100}{\milli\second}.
At this point the blue cavity focus is shifted so that the relative X-rays power turns $0$ and, at the same time, red cavity focus is also shifted so that it overlaps to the electrons. This configuration is maintained for \SI{100}{\milli\second}, then red cavity focus is shifted to turn X-rays power to $0$.
It is clear that for this technology to work, every cavity has to be stabilized during the shift, the power stored in them has to maintain almost constant before and after the movement and the movements has to be sufficiently fast, in particular less than \SI{100}{\milli\second}, so that images produced by dual color technique refers to the same configuration of the patient.

\section{Results}
\begin{figure}[ht]
\centering
\includegraphics[width=0.235\textwidth]{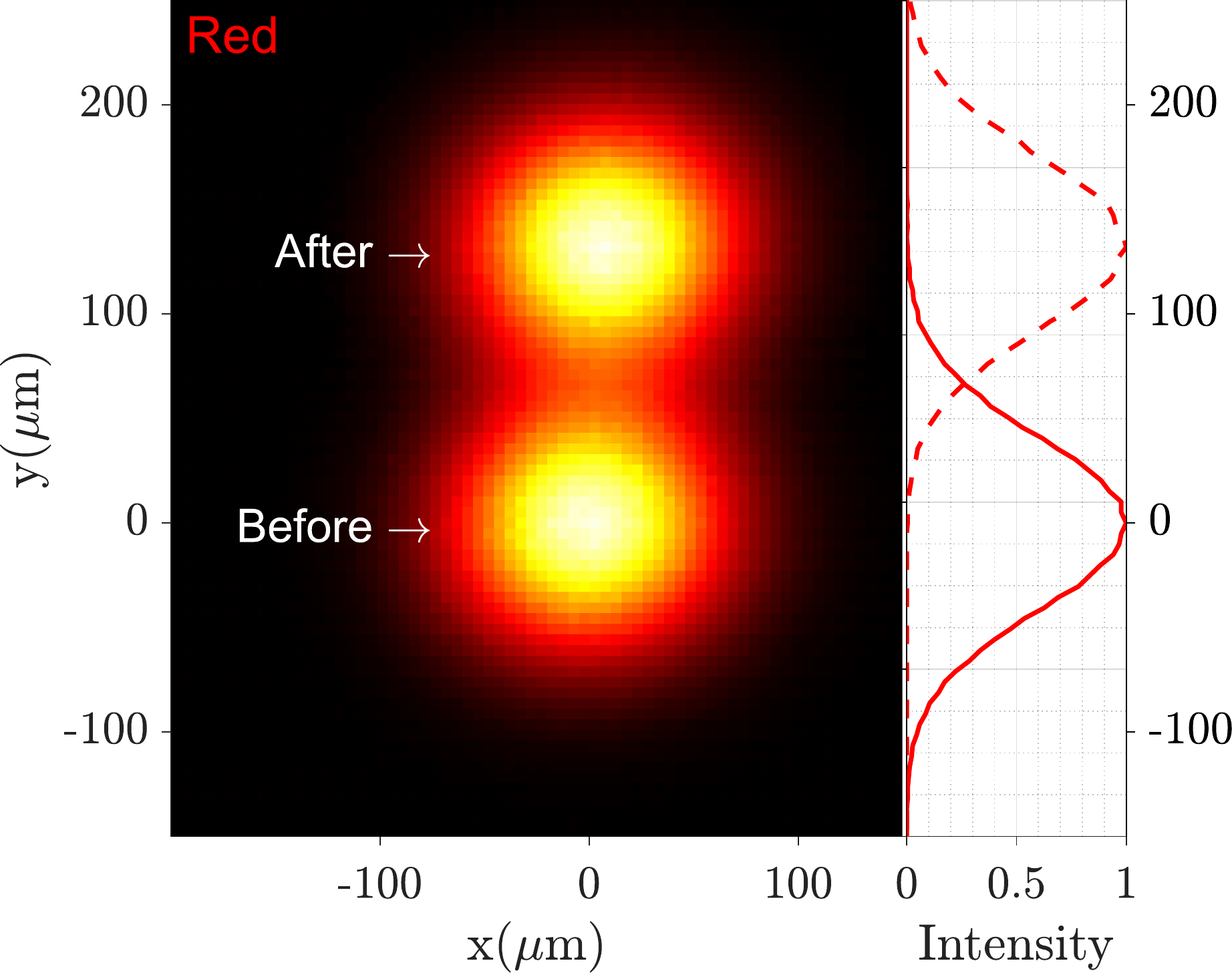}
\includegraphics[width=0.235\textwidth]{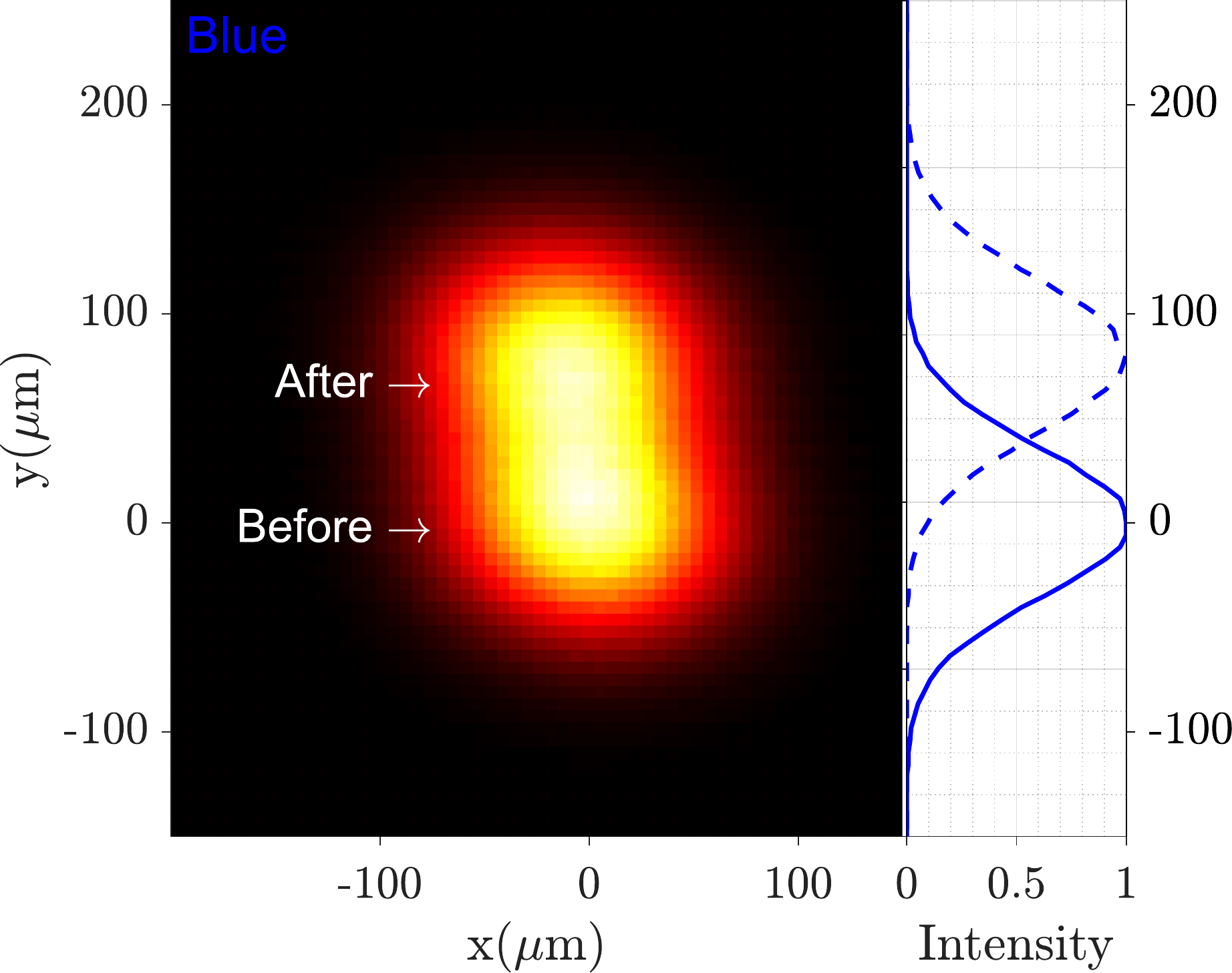}
\caption{
Imaging of the foci of the two cavities: red cavity is shown on the left, while blue cavity on the right. The images show the movement of both foci while the cavities were stabilized.
Next to each image the normalized profile of the beam for $x=\SI{0}{\micro\meter}$ is shown, where, for each cavity, the straight line indicates the profile before the movement and the dashed line after the movement.
The red cavity focus moves of about \SI{135}{\micro\meter}, while the blue cavity focus of about \SI{77}{\micro\meter}.
}
\label{fig:imaging-fuochi}
\end{figure}

Exploiting the setup described in the previous section, with the help of the pellicle beam splitter we monitored the two foci shift obtained by rotating the curved mirrors together and the result is shown in Fig.~\ref{fig:imaging-fuochi}.
Red cavity focus is shown on the left before and after the movement, with the corresponding projection for $x=\SI{0}{\micro\meter}$, while blue cavity focus is shown on the right, with the corresponding projection.
The focus of the red cavity, which is equipped with gimbal mountings, moves of about \SI{135}{\micro\meter}, while the focus of the blue cavity, equipped with standard mountings, moves a smaller distance of about \SI{77}{\micro\meter}.
The shift is greater for the red cavity because we could maintain the cavity stabilized for a greater rotation angle with gimbal mountings, since in this case the rotation just slightly affects the cavity length and the stabilization system can compensate.
Conversely, a rotation of the standard mountings greatly affects the cavity length, so it is limited by the dynamics of the stabilization system: a greater rotation causes a loss of the cavity lock to the external laser.
It is clear that the rotation must be calibrated, so that the amplitude and the timing is the same for the two curved mirrors. The calibration technique will be deepened later.
Notice that the foci dimensions are in very good agreement with the expected values calculated from the cavities geometries.
In our case the curved mirrors distance for each cavity was around \SI{4}{\milli\meter} from the confocal configuration, leading to an horizontal waist of the focus of $w_{\rm H}=\SI{92}{\micro\meter}$ and a vertical one of $w_{\rm V}=\SI{82}{\micro\meter}$. The measured dimensions was around $w_{\rm H}=\SI{93}{\micro\meter}$ and $w_{\rm V}=\SI{84}{\micro\meter}$.

\begin{figure}[ht]
\centering
\includegraphics[height=0.15\textheight]{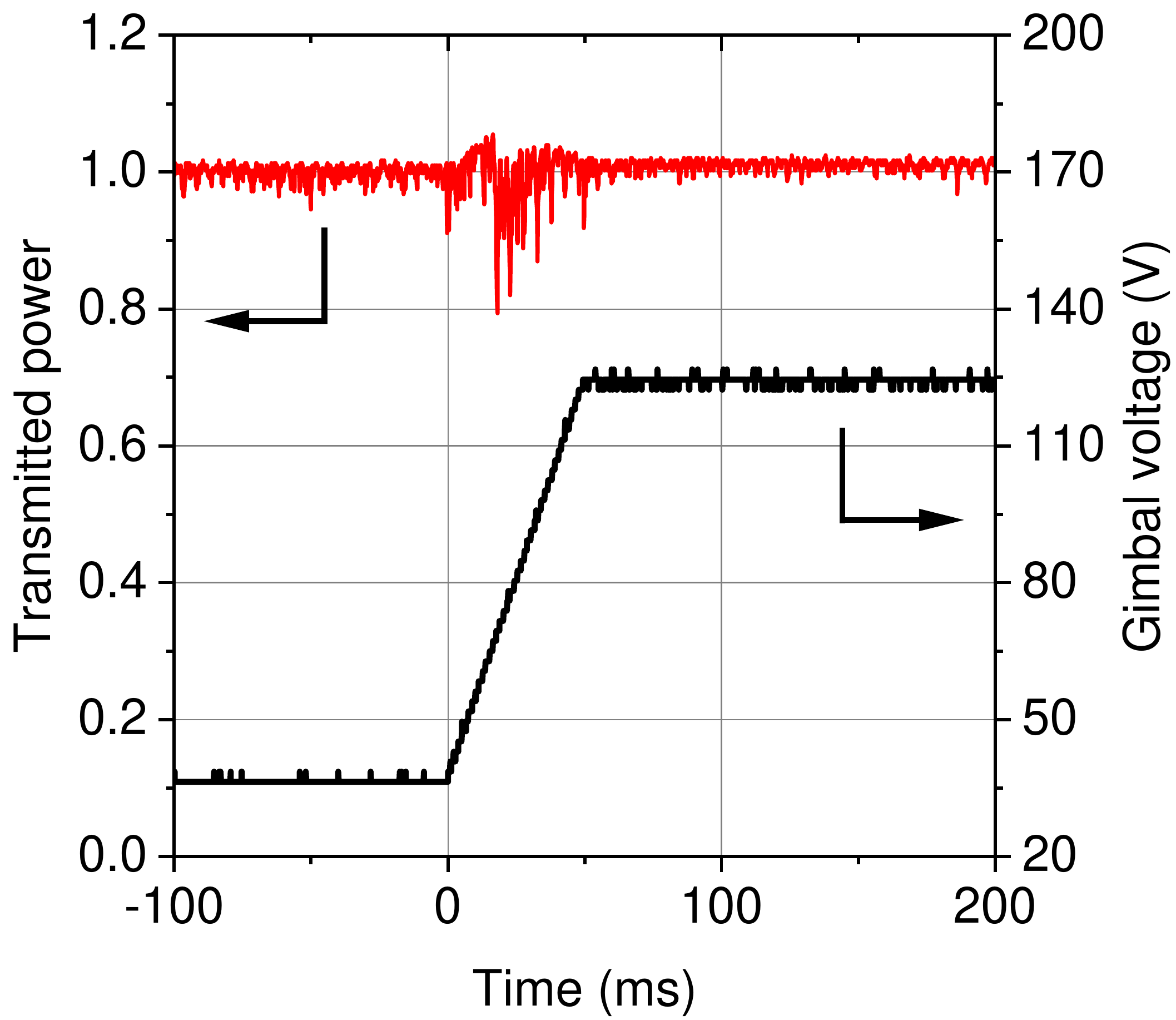} 
\includegraphics[height=0.15\textheight]{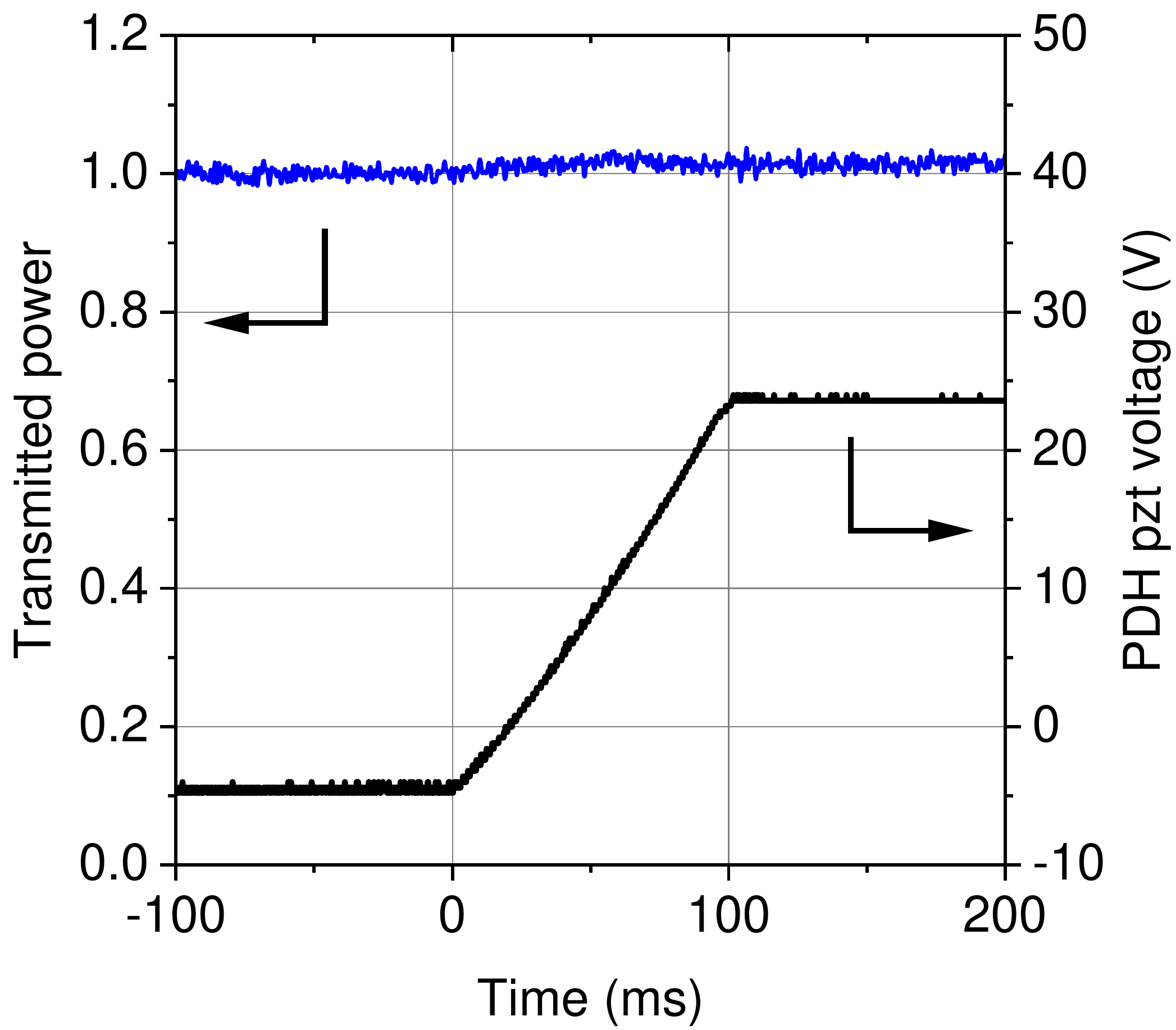}
\caption{
Normalized transmitted power (color, left axis) and movement monitor (black, right axis) for the red (left box) and blue (right box) cavities.
The movement for the red cavity is monitored by measuring the voltage applied on gimbal actuator of mirror C, while for the blue cavity by measuring the voltage on the PDH piezoelectric actuator, since the voltage on the SmarAct\textsuperscript{\textregistered} mirror is inaccessible.
The variation of the transmitted powers is \SI{1.0}{\percent} for the red cavity and \SI{1.4}{\percent} for the blue cavity. The corresponding foci movements are \SI{135}{\micro\meter} in \SI{50}{\milli\second} and \SI{77}{\micro\meter} in \SI{100}{\milli\second}, respectively.
Notice that the movements of the two foci are synchronous.}
\label{fig:trasmesso-piezo}
\end{figure}

Now that the shift is calibrated, the pellicle beam splitter is removed and the Finesse is maximum.
In this configuration we measured the transmitted powers inside the cavities during the movement, which are a monitor of the powers stored in them, and the voltages applied to the rotating actuators.
Results are shown in Fig. ~\ref{fig:trasmesso-piezo}.
The transmitted beam and the voltage on the rotating mirror piezoeletric of the red cavity is shown on the left, while of the blue cavity on the right. The voltages applied to the rotating mirrors are the same as the ones applied with the pellicle beam splitter, so the shifts of the foci is the same in the two cases.
We have a shift of the focus of \SI{0.83}{\micro\meter\per\volt} for the blue cavity, while \SI{1.54}{\micro\meter\per\volt} for the red cavity.
A SmarAct\textsuperscript{\textregistered} controller, compatible with SmarAct\textsuperscript{\textregistered} mountings, is used to rotate the blue cavity mirrors, while a National Instruments\textsuperscript{\textregistered} board connected to a high voltage driver is used to rotate the red cavity mirrors.
Everything is controlled via computer by a Labview\textsuperscript{\textregistered} program.
Each transmitted beam is detected using a power meter placed outside one of the cavity mirrors.
The variation of the transmitted beam is very low in both cases, in particular \SI{1.4}{\percent} for the blu cavity and \SI{1.0}{\percent} for the red cavity, with a very fast movement: \SI{100}{\milli\second} and \SI{50}{\milli\second}, respectively.
During the shifts, both cavities stay stabilized, so our technique fullfill the requested parameters for the dual color application.
It is worth noting that the little oscillations in the transmitted power during the movements are related to resonance frequencies of the mirrors mountings and they do not compromise the stability of the cavities.

\begin{figure}[ht]
\centering
\includegraphics[width=0.35\textwidth]{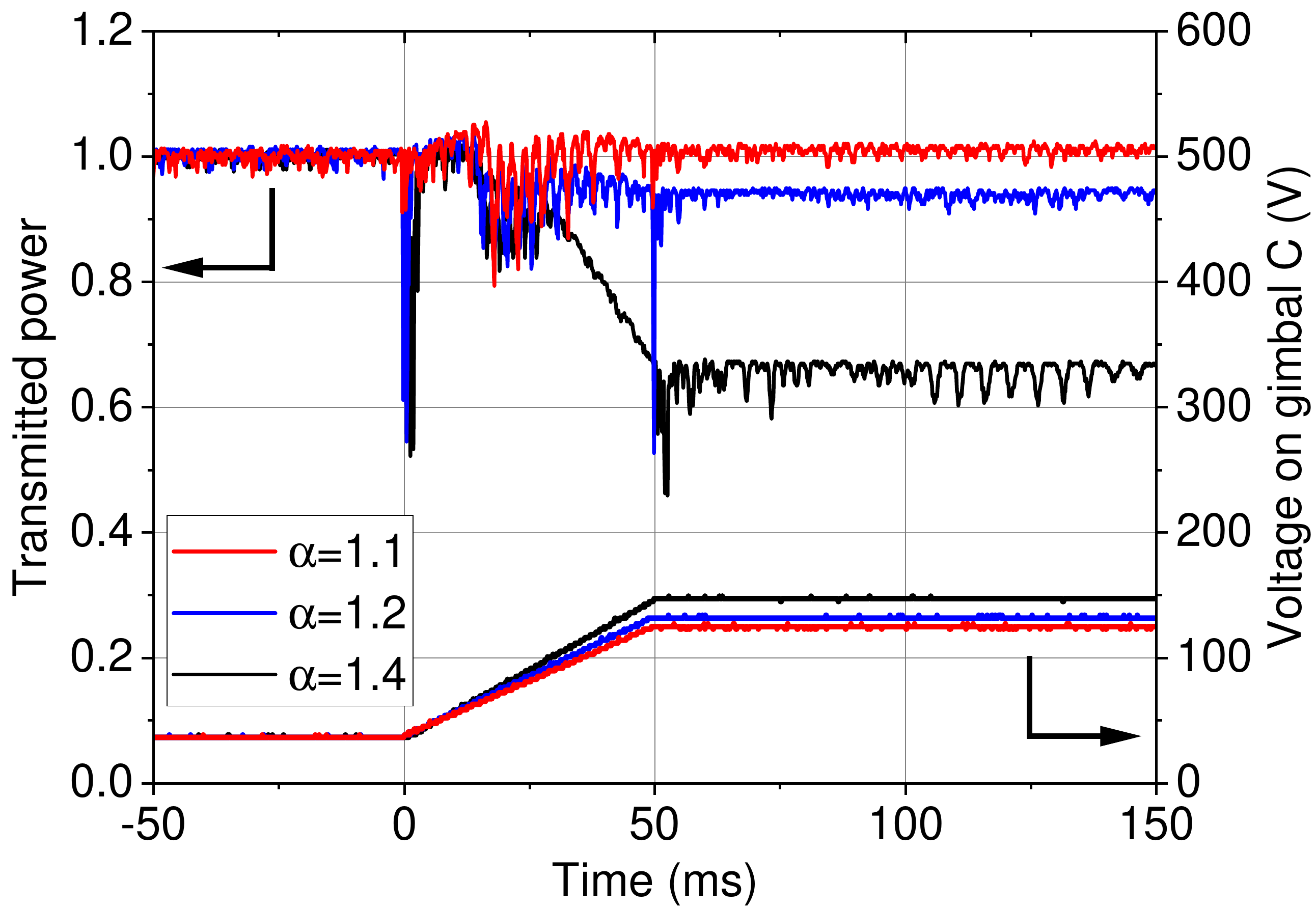}
\caption{Normalized transmitted beam of the red cavity (equipped with gimbal mountings) during a focus shift for different values of the optimization parameter $\alpha$. Best result is for $\alpha=1.1$.
Notice that if $\alpha$ is not optimized, then the cavity becomes very disaligned, resulting in a great instability.
Here the voltage on gimbal mounting of mirror B is always the same, while we change the voltage on gimbal mounting of mirror C.
}
\label{fig:ottimizzazione}
\end{figure}

As cited previously, the rotation of the two curved mirrors of each cavity must be calibrated in order to minimize the optical power variation before and after the movement.
In particular, the rotation must be synchronous and of the same amplitude. Synchronism can be easly achieved via the Labview\textsuperscript{\textregistered} program reaching the \SI{10}{\micro\second} precision, while the amplitude requires something more. Indeed, the voltage-to-stroke relation is different from one piezoeletric to another: even if it is small, it dramatically affects the aligment of our cavity, resulting in a strong variation of the power before and after the movement. The difference can be also imputed to different preloads of the two piezoelectrics.
In order to calibrate the two piezos of each cavity, we set the same starting voltage $V_{\rm 0}$, we align the cavity and then we change the voltage of a quantity $\Delta V$ for the first mirror and $\alpha \Delta V$ to the second, where $\alpha$ is our optimization parameter. $\alpha$ can be easily and fastly set and changed with our Labview\textsuperscript{\textregistered} program.
It is worth noting that we apply a starting voltage $V_{\rm 0}>\SI{10}{\volt}$ on the piezoelectrics, since around $V=0$ they present a non-linear zone in the voltage-to-stroke relation.
In Fig.~\ref{fig:ottimizzazione} we show how the power changes for different optimization parameters $\alpha$.
The best results is for $\alpha=1.1$, where the variation of the transmitted beam of \SI{1.0}{\percent}.
The non-null variation of the power can be imputed mostly to the coupling of the internal and external mode, as shown in Fig.~\ref{fig:method-sim}, but also to the non-perfect horizontal alignment: indeed, if the beam does not impinge at the same point of the curved mirrors in the horizontal direction, it results in a non-compensed asymmetry between the two mirrors during the movement.
A piezoeletric horizontal movimentation on the mountings would probably further reduce the power variation.
Notice that this calibration is fundamental for the cavity stability: when the voltage variation applied to the two piezoelectric actuators is not optimized, then the cavity becomes very disaligned, resulting in a great instability as one can see from Fig.~\ref{fig:ottimizzazione}, black curve.

\section{Conclusions}
In this work we described, experimentally implemented and validated a new optical technique to precisely overlap a laser pulse over an electron bunch in the focal point of an optical cavity while maintaining it stabilized and to
implement a dual-color method performing a synchronous and controlled movement of the foci of two high finesse actively stabilized crossed cavities.
In particular we demonstrate the feasibility of the last and more critical part, achieving a movement of the focus of \SI{135}{\micro\meter} in a time of \SI{50}{\milli\second}. The cavity was stabilized during the movement and the variation in the stored power before and after the movement was negligible for our application (\SI{1.0}{\percent}).
Moreover, we explained a simple technique used to optimize the variation of the optical power before and after the movement and another one for the imaging of the foci by means of pellicle beam splitter.

Our method seems to be promising to implement in ICS system the generation of dual-color X-rays, since it allows the fine adjustment of the focal point position in a actively stabilized high finesse optical cavity without the necessity of moving the entire optical table with respect to the electron bunch.

\bibliography{biblio}

\end{document}